\documentclass[sigconf]{acmart}
\AtBeginDocument{%
  }

\usepackage{multirow}
\usepackage{caption}
\usepackage{tcolorbox}
\usepackage{multicol}

\copyrightyear{2025}
\acmYear{2025}
\setcopyright{rightsretained}
\acmConference[SIGCSE TS 2025]{Proceedings of the 56th ACM Technical Symposium on Computer Science Education V. 2}{February 26-March 1, 2025}{Pittsburgh, PA, USA}
\acmBooktitle{Proceedings of the 56th ACM Technical Symposium on Computer Science Education V. 2 (SIGCSE TS 2025), February 26-March 1, 2025, Pittsburgh, PA, USA}
\acmDOI{10.1145/3641555.3705189}
\acmISBN{979-8-4007-0532-8/25/02}

\begin{document}

\title{Generating AI Literacy MCQs: A Multi-Agent LLM Approach}

\author{Jiayi Wang}
\email{jiayiwang2025@u.northwestern.edu}
\orcid{0009-0003-1665-9360}
\affiliation{%
  \institution{Northwestern University}
  \city{Evanston}
  \state{Illinois}
  \country{USA}
}

\author{Ruiwei Xiao}
\email{ruiweix@andrew.cmu.edu}
\orcid{0000-0002-6461-7611}
\affiliation{%
  \institution{Carnegie Mellon University}
  \city{Pittsburgh}
  \state{Pennsylvania}
  \country{USA}
}

\author{Ying-Jui Tseng}
\email{yingjuit@andrew.cmu.edu}
\orcid{0009-0006-1801-6061}
\affiliation{%
  \institution{Carnegie Mellon University}
  \city{Pittsburgh}
  \state{Pennsylvania}
  \country{USA}
}


\begin{abstract}
Artificial intelligence (AI) is transforming society, making it crucial to prepare the next generation through AI literacy in K-12 education. However, scalable and reliable AI literacy materials and assessment resources are lacking. To address this gap, our study presents a novel approach to generating multiple-choice questions (MCQs) for AI literacy assessments. Our method utilizes large language models (LLMs) to automatically generate scalable, high-quality assessment questions. These questions align with user-provided learning objectives, grade levels, and Bloom's Taxonomy levels. We introduce an iterative workflow incorporating LLM-powered critique agents to ensure the generated questions meet pedagogical standards. In the preliminary evaluation, experts expressed strong interest in using the LLM-generated MCQs, indicating that this system could enrich existing AI literacy materials and provide a valuable addition to the toolkit of K-12 educators.
\end{abstract}


\maketitle

\section{Introduction}

AI is rapidly transforming our world, from the way we work to the way we interact with each other. To prepare the next generation for this AI-driven world, educational frameworks like AI4K12's "Five Big Ideas"\cite{Touretzky2022MachineLA} (Perception, Representation and Reasoning, Learning, Natural Interaction, and Natural Interaction) have been developed to integrate AI literacy into K-12 education. 

However, as a relatively new subject area, there is a lack of learning materials in AI literacy instructional and assessment activities\cite{gardner2019ai}. In addition, existing courses and assessment resources are often not scalable or reliable \cite{tseng2024assessing}. This shortage presents a significant challenge for teachers who must teach and assess students' understanding of AI concepts while fostering critical thinking skills for responsible engagement with these technologies.

Leveraging Large Language Models (LLMs) to generate assessment questions \cite{doughty2024comparative,caines2023application,stamper2024enhancing} offers a potential solution. This method has shown promise in subjects like programming\cite{doughty2024comparative}, medicine\cite{indran2024twelve}, and language learning\cite{caines2023application}. The potential to apply these methods to AI literacy assessments could help bridge the gap for educators by providing accessible, scalable assessment resources.

This exploratory study, conducted as part of the active.ai\footnote{https://active-ai.vercel.app} project, builds on previous research on MCQ generation \cite{doughty2024comparative, moore2023assessing} and AI literacy \cite{Touretzky2022MachineLA}. It seeks to address the following research question: \textbf{Can LLM-powered multi-agent workflows effectively generate high-quality MCQs for AI literacy?}

\section{Methodology}

\begin{figure}[h]
  \centering
  \includegraphics[width=\linewidth]{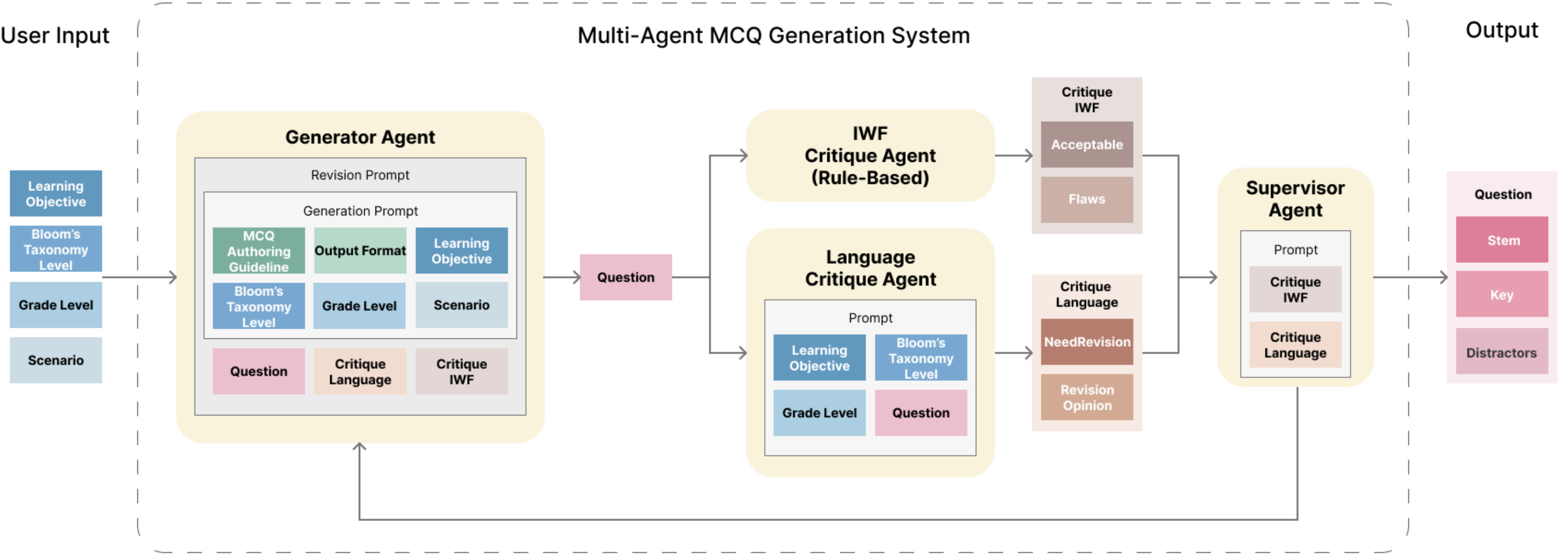}
  \caption{Multi-Agent MCQ Generation System.}
\end{figure}


The Multi-Agent MCQ Generation System developed for this study generates and refines MCQs to assess AI literacy, using a workflow built on the LangGraph framework\footnote{https://www.langchain.com/langgraph}, and OpenAI's \texttt{gpt-4o-mini-2024-07-18} model.

User inputs, including learning objectives, Bloom’s taxonomy levels \cite{andersonTaxonomyLearningTeaching2001}, grade level, and specific scenarios, guide the Generator Agent in producing an initial question with fields \textit{stem} (question), \textit{key} (correct option), and \textit{distractors} (incorrect options). The question is independently reviewed by two critique agents: a Language Critique Agent, which evaluates readability and alignment with grade level, and an IWF (Item-Writing Flaw) Critique Agent \cite{moore2023assessing}, which applies rule-based checks (modified from \cite{moore2023assessing}) for issues such as implausible distractors or absolute terms. A question with 0 or 1 flaw is considered acceptable. Based on feedback from these agents, the question is either approved by the Supervisor Agent or sent back to the Generator Agent for revision. This iterative process continues until the question meets quality standards or reaches the maximum number of revisions. We generated a total of 40 multiple choice questions for students in grades K7-9. 

The following is an example of a generated question:
\vspace{0pt}

\scalebox{0.85}{
\begin{tcolorbox}[title=Sample Generated Question, fontupper=\footnotesize]
Ben is considering using an AI tool to help him write a creative story. Which of the following reasons best explains when using AI might be a bad choice for his learning?	

A. It may produce a story that lacks originality and personal expression.

B. AI can provide quick feedback on grammar and structure.	

C. Using AI can help him brainstorm new ideas for his story.	

D. AI tools can assist in organizing his thoughts more effectively.
\end{tcolorbox}
}



\section{Results}

Three experts with more than a year of K-12 AI Literacy teaching experiences, evaluated the quality of each of the 40 MCQs used a ten-item rubric (Table \ref{tab:rubric} modified from \cite{scaria2024automated}). 

\begin{table}[h]
  \fontsize{8pt}{9pt}\selectfont 
  
  \scalebox{0.78}{
  \begin{tabular}{lp{\dimexpr 0.45\textwidth-2\tabcolsep}}
    \toprule
    \textbf{Rubric Item} & \textbf{Definition and Response Option} \\
    \midrule
    \textbf{Understandable} & Could you understand what the question is asking? \textit{(yes/no)} \\
    \textbf{LORelated} & Is the question related to the learning objective? \textit{(yes/no)} \\
    \textbf{Grammatical} & Is the question grammatically well-formed? \textit{(yes/no)} \\
    \textbf{Clear} & Is it clear what the question asks for? \textit{(yes/more\_or\_less/no)} \\
    \textbf{Rephrase} & Could you rephrase the question to make it clearer and error-free? \textit{(yes/no)} \\
    \textbf{Answerable} & Can students answer the question with the information or context provided within? \textit{(yes/no)} \\
    \textbf{Central} & Do you think being able to answer the question is important to work on the topic given in the prompt? \textit{(yes/no)} \\
    \textbf{WouldYouUseIt} & If you were a teacher teaching the course topic would you use this question or the rephrased version in the course? \textit{(this/rephrased/both/neither)} \\
    \textbf{Bloom'sLevel} & Do you think the question is of the Bloom's taxonomy level labeled? \textit{(yes/no)} \\
    \textbf{GradeLevel} & Do you think the question is appropriate for K7-9? \textit{(yes/no)} \\
    \bottomrule
\end{tabular} 
}
\caption{Rubric for Question Evaluation}
  \label{tab:rubric}
\end{table}



\begin{figure}[b]
\vspace{-3mm}
  \centering
  \includegraphics[width=\linewidth]{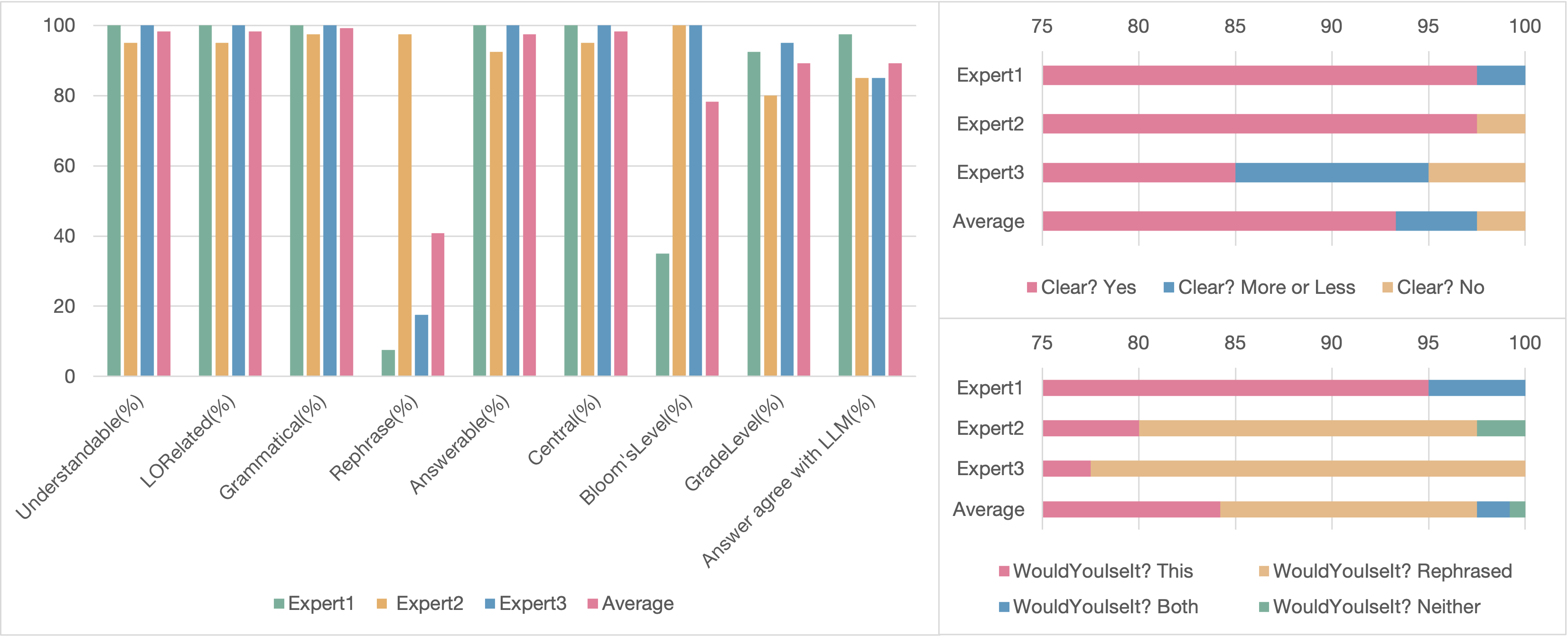}
  \caption{Evaluation Result}
\vspace{-3mm}
\end{figure}

Experts 1, 2, and 3 agreed with the system on 97.5\%, 85\%, and 85\% of the correct answers, respectively. This suggests that the system generally produces correct answers that align with expert judgments. Although rare, there are also occasional misalignments between the system’s generated answers and expert evaluations.

There was strong agreement among experts on syntactical correctness, with high "Yes" responses for criteria such as \textit{Understandable}, \textit{Answerable}, \textit{Grammatical}, and \textit{Clear}, with average percentages ranging from 93.3\% to 99.2\%. This suggests that the questions generated by the multi-agent approach are well-formed and clear. Additionally, experts agreed that the questions were highly relevant to the learning objectives (\textit{LORelated} 98.3\%) and central to the topics being assessed (\textit{Central} 98.3\%).

There was noticeable disagreement among the experts in criterion \textit{Rephrase}, \textit{Bloom’sLevel}, and \textit{GradeLevel}. Expert 2 was stricter, suggesting that 97.5\% of the questions needed rephrasing, while Experts 1 and 3 flagged fewer questions (7.5\% and 17.5\%). The same trend is seen with the \textit{GradeLevel}, where Expert 2 marked 20\% of the questions as inappropriate, compared to Expert 1 at 7.5\% and Expert 3 at 5\%. These discrepancies likely arise from varying interpretations of the rubric or differences in pedagogical approaches. Expert 1 significantly differed from Experts 2 and 3 on the \textit{Bloom'sLevel} criterion, marking only 35\% of the questions as appropriate for the intended cognitive level, while the other two experts rated all questions as aligned with the correct Bloom’s level. 

Finally, there was moderate agreement on the \textit{WouldYouUseIt} criterion, with expert responses ranging from 77.5\% to 95\% (average 84.2\%). This indicates that a significant majority of the questions were seen as suitable for classroom use, demonstrating that the multi-agent system can produce questions that are not only well-formed but also useful in practical educational settings. However, the diverse responses suggest that individual preferences or teaching styles influence the willingness to use certain questions.

Overall, while some refinement is necessary to accommodate diverse educational contexts, the multi-agent approach demonstrates strong potential to generate pedagogically sound and scalable high-quality assessment questions for AI literacy in K-12 education.

\section{Limitations and Future Work}

One avenue for future work is to conduct classroom trials to evaluate student learning outcomes and teacher usability. While our multi-agent system successfully generated AI literacy MCQs, it has not been tested in real-world classrooms, so its effectiveness in diverse learning environments is still uncertain. Another important direction for future research is to explore additional methods for evaluating question quality beyond expert assessments, such as leveraging student performance data, crowd-sourced feedback, and automated analysis to provide a more comprehensive and objective evaluation. Our current evaluation relied on expert judgments, which introduces potential subjectivity. Key priorities include expanding the system’s adaptability to more grade levels, Bloom’s Taxonomy levels, and subjects beyond AI literacy, as well as integrating diverse assessment formats (e.g., open-ended, project-based questions) to enhance its practical utility. Finally, the tool can help teachers design AI literacy courses using a backward design approach, aligning assessments with learning objectives.


\bibliographystyle{ACM-Reference-Format}
\bibliography{reference}

\end{document}